%% file: main_deteQt.tex
\documentclass[aps,twocolumn,prx]{revtex4-2}

\usepackage{bm}
\usepackage{bbm}
\usepackage{bbold}
\usepackage{amsfonts}
\usepackage{amsmath}
\usepackage{amssymb}
\usepackage{graphicx}
\usepackage{mathtools}
\usepackage{upgreek}
\usepackage{complexity}
\usepackage{xfrac}
\usepackage{soul}
\usepackage{xcolor}
\usepackage{dsfont}
\usepackage{tikz}
\usetikzlibrary{quantikz2}
\usepackage{titlesec}
\usepackage{comment}

\usepackage[colorlinks=true,citecolor=blue]{hyperref}
\hypersetup{colorlinks=true,citecolor=blue,linkcolor=blue,urlcolor=blue}
\usepackage{url}

\let\oldbibitem\bibitem % copy bibitem into oldbibitem
\renewcommand{\bibitem}{
    \renewcommand{\doi}[1]{\texttt{\href{https://doi.org/##1}{doi:##1}}} % override \doi
    \let\bibitem\oldbibitem % restore \bibitem
    \oldbibitem % call old \bibitem
}

\newcommand{\exeter}{Department of Physics and Astronomy, University of Exeter, Stocker Road, Exeter EX4 4QL, United Kingdom}

\begin{document}

\title{Quantum community detection via deterministic elimination}

\author{Chukwudubem Umeano}
\affiliation{\exeter}
\author{Stefano Scali}
\affiliation{\exeter}
\author{Oleksandr Kyriienko}
\affiliation{\exeter}

\begin{abstract}
We propose a quantum algorithm for calculating the structural properties of complex networks and graphs. The corresponding protocol ---\emph{deteQt} --- is designed to perform large-scale community and botnet detection, where a specific subgraph of a larger graph is identified based on its properties. We construct a workflow relying on ground state preparation of the network modularity matrix or graph Laplacian. The corresponding maximum modularity vector is encoded into a $\log(N)$-qubit register that contains community information. We develop a strategy for ``signing'' this vector via quantum signal processing, such that it closely resembles a hypergraph state, and project it onto a suitable linear combination of such states to detect botnets. As part of the workflow, and of potential independent interest, we present a readout technique that allows filtering out the incorrect solutions deterministically. This can reduce the scaling for the number of samples from exponential to polynomial. The approach serves as a building block for graph analysis with quantum speed up and enables the cybersecurity of large-scale networks.
\end{abstract}

\maketitle
% \tableofcontents

%%%%%%%%%%%%%%%%%%%%%%%%%%%%
%%%%%%%%%%%%%%%%%%%%%%%%%%%%
%%%%%%%%%%%%%%%%%%%%%%%%%%%%

\section*{Introduction}
\label{sec:intro}

Quantum computing (QC) has emerged as a computing paradigm that suits problems with a specific structure \cite{Aaronson2013book}. Its applications include problems with inherently quantum behavior, with notable examples lying within chemistry \cite{McArdle2020RMP} and strongly-correlated materials \cite{ALEXEEV2024}, or problems with an algebraic structure \cite{Simon1997,Aaronson2010}. Here, the prominent example is an exponential speedup via Shor's algorithm for factoring large integers~\cite{Shor_1997}. This is an example of a $\BQP$ complexity class algorithm that relies on efficient solution readout where only a few samples are required (logarithmic in the input size). Other domains that can benefit from quantum subroutines include linear equation solvers \cite{harrow2009quantum,childs2017quantum}, matrix inversion \cite{Lin2020optimalpolynomial,Subasi2019,chakraborty2024implementing}, and differential equation solvers \cite{Costa2022,williams2024jacobi}. However, here the advantage largely depends on the ability to read the solution \cite{Linden2020}, representing a challenge \cite{Biamonte2017}. Finally, quantum computing offers advantage in sampling quasi-probability distributions \cite{Arute2019,Hangleiter2023}, motivating the development of quantum generative modeling \cite{Zoufal2019,Coyle2020,Paine2023QQM,Kyriienko2024DQGM,kasture2022protocols,Wu2024}.

Moving away from natively quantum or algebraic problems, there remains an open question if quantum computers can help solve problems with a combinatorial structure. On one hand, for provably $\NP$-hard instances of optimization problems we do not expect an exponential advantage \cite{Abbas2024}, while there is still room for efficient heuristics \cite{LeoZhou2020,Ebadi2022,ZhuEconomou2022}. However, recent advances show a largely improved ability to approximate certain optimization problems via decoded quantum interferometry \cite{Jordan2024DQI}, triggering further search. Similarly, quantum computers shown promise in topological data analysis (TDA)~\cite{Edelsbrunner_2002, Zomorodian_2004, Carlsson_2009, Wasserman_2018, Chazal_2021}, where quantum algorithms can provide insights that are difficult to achieve with classical methods~\cite{Lloyd2016, Gyurik_2022, Berry_2022, Scali2024_1, Scali2024_2, akhalwaya2024tda}. In the domain of graph and network theory, there are examples of quantum algorithms for finding shortest paths~\cite{D_rr_2006, Krauss_2020} and detecting cliques~\cite{Chapuis_2017}. Photonic samplers were successfully applied to study maximal clique finding for molecular docking \cite{Arrazola2018,Banchi2020,Yu2023}. The exponential improvement for quantum walks on glued trees suggests a further advantage for selected graph-based problems \cite{Childs2003}. Interestingly, there are also $\BQP$ algorithms with exponential advantage in simulating large networks of coupled oscillators \cite{Babbush2023}.  

Motivated by the examples above, we consider a wide area of complex networks~\cite{Strogatz_2001} as a potential beneficiary of quantum computing. Complex networks (CNs) represent large multi-node graphs with connectivity defined by relations between objects that they model. CNs are ubiquitous in various domains, representing intricate systems such as communication networks, biological networks, social networks, and financial systems~\cite{Albert_2002, BOCCALETTI_2006, Bardoscia_2021}. One of the key tasks in analyzing these networks is community detection \cite{Fortunato_2010}, where the objective is to identify groups of nodes that are more interconnected relative to the rest of the network. This task is crucial in various applications, including social media analysis, where communities may represent groups with shared interests~\cite{Liao_2017}; biology, where communities might correspond to functional modules in a protein interaction network~\cite{de_Silva_2005, Deeds_2012}; and cybersecurity, where identifying clusters of compromised nodes can help in the detection of botnets or coordinated cyber-attacks~\cite{He_2022}. Despite its importance, community detection becomes increasingly challenging as networks grow in size and complexity~\cite{Klamt_2002}. To date, the attempts to connect quantum systems and complex networks mostly remained at a conceptual level \cite{Biamonte2019} and much work is needed to bring full QC benefits to hard community detection and graph partitioning problems. 
%%%
\begin{figure*}[t!]
    \centering
    \includegraphics[width=\linewidth]{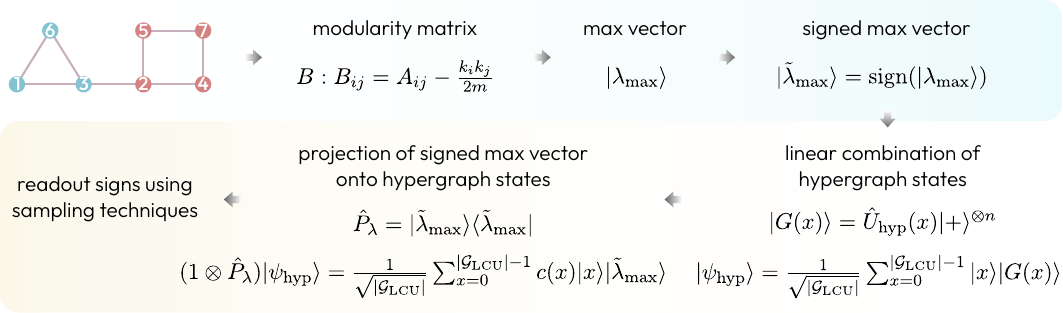}
    \caption{Flowchart summarizing \emph{deteQt} --- the developed quantum algorithms for modularity-based community and botnet detection. The protocol consists of several steps, including finding the leading eigenvector of the modularity matrix, and readout sampling techniques to infer its signs. We detail each step and subroutine in Sec.~\ref{sec:protocol}.}
    \label{fig:protocol}
\end{figure*}
%%%

Classical algorithms for community detection, such as the Girvan-Newman algorithm~\cite{Newman_2004}, modularity-based spectral methods~\cite{Newman_2006}, and spectral clustering~\cite{von_Luxburg_2007}, have made significant strides in this area. However, these methods often struggle with scalability and accuracy in large-scale networks. The graph partitioning problem~\cite{Ku_era_1995, Bulu__2016}, closely related to community detection, is also of significant interest, particularly in the context of optimizing the division of a network into smaller, more manageable subgraphs~\cite{Andreev_2004}. This problem is critical in parallel computing~\cite{Hendrickson_2000}, VLSI design~\cite{Kahng_2022}, and load balancing~\cite{Patil_2021}. Current state-of-the-art classical protocols, like the Louvain method~\cite{De_Meo_2011} and the use of Markov Clustering (MCL)~\cite{Van_Dongen_2008}, attempt to partition large networks efficiently but are limited by computational resources and the exponential growth of problem complexity.

%Quantum computers offer a promising avenue for addressing these challenges, particularly in analyzing the topological properties of networks~\cite{Strogatz_2001}. %This potential is closely related to topological data analysis (TDA)~\cite{Edelsbrunner_2002, Zomorodian_2004, Carlsson_2009, Wasserman_2018, Chazal_2021}, where quantum algorithms can provide insights that are difficult to achieve with classical methods~\cite{Lloyd2016, Gyurik_2022, Scali2024_1, Scali2024_2}. However, there remains a gap in the application of quantum methods to the structural and topological analysis of complex networks, particularly in fields like cybersecurity, where early and accurate detection of malicious activity is critical.

In this paper, we propose a quantum algorithm designed to calculate the structural properties of complex networks and graphs. Our protocol, named \emph{deteQt}, is specifically developed for large-scale community and botnet detection. We introduce the workflow based on ground state preparation for a modularity matrix and quantum spectral analysis. By utilizing tools such as block-encodings, linear combinations of unitaries, and quantum singular value transformation, we project the leading eigenvector of the modularity matrix, which encodes the community structure, to identify distinct communities. Furthermore, we introduce an adaptive readout technique that efficiently identifies nodes under specified conditions, which we coin as deterministic elimination (defined by properties of hypergraph states). 

Looking toward applying quantum methods to the structural and topological analysis of complex networks, we consider the field of cybersecurity, where early and accurate detection of malicious activity is critical. We connect our protocols with malicious botnet detection, where the structure of such a subgraph is efficiently discovered. The proposed \emph{deteQt} approach not only provides a building block for quantum-accelerated graph analysis but also enhances the cybersecurity of large-scale networks. %This has significant implications for cryptography~\cite{Rivest_1978, Pirandola_2020}.

%%%%%%%%%%%%%%%%%%%%%%%%%%%%
%%%%%%%%%%%%%%%%%%%%%%%%%%%%
%%%%%%%%%%%%%%%%%%%%%%%%%%%%

\section{Protocol: subroutines}
\label{sec:protocol}

We proceed to describe the quantum approaches to studying complex network properties, in particular concentrating on graph partitioning, community structure learning, and botnet detection. To this end, we have developed a quantum protocol to tackle the problem of modularity-based community detection. We refer to it as the \emph{deteQt} protocol, which can use an exponentially-improved number of operations as compared to classical analogues, for specified readout conditions and network types. 
A schematic of the full \emph{deteQt} protocol is given in Fig.~\ref{fig:protocol}, where we describe the required steps from instantiating the network to sampling the partitions. 

\subsection{Input}

Our procedure is able to work on any network. However, it is most effective when there is a clear community structure; i.e. there exists a subset of nodes that are densely connected to each other, but sparsely connected to the rest of the network. We encode the graph using the modularity matrix $B$, with elements
\begin{equation}
    B_{ij} = A_{ij} - \frac{k_ik_j}{2m},
\end{equation}
where $A_{ij}$ are elements of the adjacency matrix: $A_{ij}=1$ if node $i$ is connected to node $j$, otherwise $A_{ij}=0$. As before, $k_i$ is the degree of node $i$, and $m$ is the total number of edges in the network.

\subsection{Ground state preparation of max vector}
\label{sec:groundstate}

The task of modularity-based community detection is to find the optimal vector $s^{\mathrm{opt}}$ that maximizes the modularity $Q$ \cite{Fortunato_2010}. Modularity-based methods can be used to divide a network into an arbitrary number of communities. Here, we focus on partitioning into two communities. In this case, we can write the equation for modularity as
\begin{equation}
    \label{eq:modularity}
    Q=\frac{1}{4m}\sum_{ij}B_{ij}s_is_j = \frac{1}{4m}\mathbf{s^TBs},
\end{equation}
where $s_i = \pm 1$ depending on which community node $i$ belongs to. Note that hierarchical approaches can be used once we need to detect more communities \cite{Fortunato_2010}. An effective way of finding $s^{\mathrm{opt}}$ is to find the eigenvector corresponding to the largest eigenvalue of $\mathbf{B}$. We refer to this leading eigenvector as the max vector, or $|\lambda_{\mathrm{max}}\rangle$. Taking the sign of the elements of $|\lambda_{\mathrm{max}}\rangle$ gives us the quasi-optimal bipartitioning $|s^{\mathrm{opt}}\rangle = \mathrm{sign}(|\lambda_{\mathrm{max}}\rangle)$. From here on, we will refer to the smaller partition as the \textit{botnet}.

Since the modularity matrix for an $N$-node network is an $N \times N$ matrix, it can be loaded onto a quantum computer using $n=\lceil \log_2 N \rceil$ qubits. To prepare $|\lambda_{\mathrm{max}}\rangle$, we change $B\rightarrow -B$, such that $|\lambda_{\mathrm{max}}\rangle$ is now the ground state. This can be efficiently prepared on a quantum computer for operators (Hamiltonians) that are strongly correlated \cite{McArdle2020RMP,Gilyen2019}. There are several quantum techniques for ground state preparation (GSP). The right choice of procedure depends on the resources available and the fidelity required.

\subsection{``Signing" the max vector}

Once we have the ground state $|\lambda_{\mathrm{max}}\rangle$, which can always be expanded in the computational basis as  $|\lambda_{\mathrm{max}}\rangle = \sum_{x=0}^{2^n-1}c_x|x\rangle$, the next task is to apply the sign function onto the amplitudes of the state. This transforms it into a real equally weighted (REW) state of the form \cite{qu2012quantum}
\begin{equation}\label{REW_state}
    |\psi_{\mathrm{REW}}^n\rangle = \frac{1}{\sqrt{2^n}}\sum_{x=0}^{2^n-1}\mathrm{sign}(c_x)|x\rangle,
\end{equation}
with $\mathrm{sign}(c_x) = 1$ if $c_x > 0$ and $\mathrm{sign}(c_x) = -1$ otherwise (note that the amplitudes of $|\lambda_{\mathrm{max}}\rangle$ are always real). For example, if we have the max vector $|\lambda_{\mathrm{max}}\rangle = [0.602, 0.372, -0.602, 0.372]$, we aim to transform it into the REW state $|\Tilde{\lambda}_{\mathrm{max}}\rangle = \frac{1}{2}[1,1,-1,1]$. REW states have a prominent role in several primordial quantum algorithms, such as Simon's \cite{Simon1997}, Grover's \cite{Grover1996} and the Deutsch-Jozsa algorithm \cite{deutsch1992rapid}. 

We have two main ingredients for performing the transformation $|\lambda_{\mathrm{max}}\rangle\rightarrow |\Tilde{\lambda}_{\mathrm{max}}\rangle$. The first step is the block encoding of the amplitudes of $|\lambda_{\mathrm{max}}\rangle$ on the diagonal of a matrix \cite{Low2019hamiltonian}. For simplicity, we give the following details assuming we have a $n$-qubit unitary-preparation circuit $\hat{U}_{\mathrm{max}}$ such that $\hat{U}_{\mathrm{max}}|0\rangle^{\otimes n} = |\lambda_{\mathrm{max}}\rangle$. This encompasses GSP strategies such as variational quantum eigensolver \cite{Cerezo2021rev} or adiabatic state preparation \cite{Subasi2019}, where $\hat{U}_{\mathrm{max}}$ would be our trained variational ansatz. However, it is important to note that this block-encoding protocol can be extended to non-unitary ground state preparation methods \cite{rattew2023non}.

By using a circuit involving calls to controlled-$\hat{U}_{\mathrm{max}}$, we can prepare a $(2n+2)$-qubit block-encoding unitary $\hat{U}_{\mathrm{BE}}$ \cite{rattew2023non} corresponding to
\begin{equation}\label{Block_enc_eq}
    \hat{U}_{\mathrm{BE}} = \begin{pmatrix} * & * \\ * & A \end{pmatrix},~ A=\left(\langle0^{\otimes n+2}| \otimes \mathds{1}_n\right)\hat{U}_{\mathrm{BE}}\left(|0^{\otimes n+2}\rangle \otimes \mathds{1}_n\right),
\end{equation}
and the matrix $A$ is a diagonal matrix containing the amplitudes of $|\lambda_{\mathrm{max}}\rangle$,
\begin{equation}
    A=\begin{pmatrix}
        c_0 & 0 & \ldots & 0 \\ 0 & c_1 & \ldots & 0 \\ \vdots & \vdots & \ddots & \vdots \\ 0 & 0 & \ldots & c_{2^n-1} 
    \end{pmatrix}.
\end{equation}
We show the circuits to perform the block encoding in Fig.~\ref{fig:U_BE}.

\begin{figure*}
    \centering
    \includegraphics[width=\linewidth]{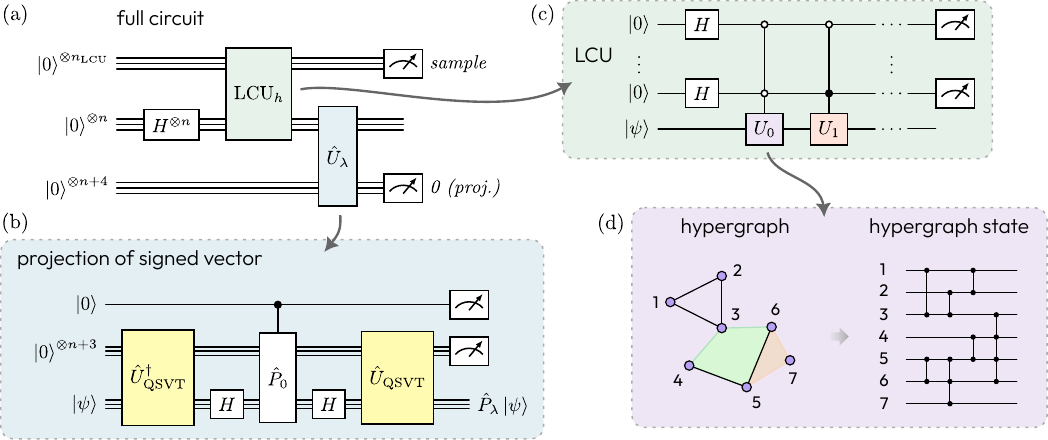}
    \caption{(a) Full circuit for projecting the signed max vector onto the hypergraph superposition state. The first layer of Hadamards prepares the state $|+\rangle^{\otimes n}$, which the hypergraph circuits act upon [as in Eq.~\eqref{hypergraph_eq}]. (b) Circuit  $\hat{U}_\lambda$ for the projection of the signed max vector onto the system register (when $\boldsymbol{0}$ is measured on the $(n+4)$ ancilla qubits). Here $\hat{P}_{0} = \ket{0}^{\otimes 2n+3}\bra{0}^{\otimes 2n+3}$ and $H \equiv H^{\otimes n}$. (c) LCU circuit for the preparation of the hypergraph superposition state. The LCU$_h$ block has the standard form of the LCU circuit but without the final PREPARE operation (in this case, a layer of Hadamards). The top register is then sampled to retrieve the bipartition. (d) Mapping between the hypergraphs and the circuits preparing the hypergraph states which make up the LCU. Black lines represent edges between the nodes of the hypergraph. These translate into CZ gates (black connected dots) between corresponding qubits in the hypergraph state generation. Shaded regions correspond to hyperedges between multiple qubits, e.g. 5-6-7 and 3-4-5-6 in the figure. These translate into multi-controlled phase gates. The multi-qubit CZ gates act on a $|+\rangle^{\otimes 7}$ state prepared from the computational zero state with Hadamards.}
    \label{fig:Full_circuit}
\end{figure*}

Now that we have the amplitudes of $|\lambda_{\mathrm{max}}\rangle$ encoded along the diagonal of a matrix, we can use quantum singular value transformation (QSVT) to apply the sign function directly onto the amplitudes. 
The full QSVT sequence $\hat{U}_{\mathrm{QSVT}}$ is a $(2n+3)$-qubit block-encoding of the transformed matrix $\mathrm{sign}(A)$ corresponding to
\begin{align}\label{u_qsvt}
    &\hat{U}_{\mathrm{QSVT}} = \begin{pmatrix} * & * \\ * & \mathrm{sign}(A) \end{pmatrix}, \\ &\mathrm{sign}(A)=\left(\langle0^{\otimes n+3}| \otimes \mathds{1}_n\right)\hat{U}_{\mathrm{QSVT}}\left(|0^{\otimes n+3}\rangle \otimes \mathds{1}_n\right),
\end{align}
and the matrix $\mathrm{sign}(A)$ is a diagonal matrix containing the signed amplitudes of $|\lambda_{\mathrm{max}}\rangle$,
\begin{align}
    &\mathrm{sign}(A)=\begin{pmatrix}
        \mathrm{sign}(c_0) & 0 & \ldots & 0 \\ 0 & \mathrm{sign}(c_1) & \ldots & 0 \\ \vdots & \vdots & \ddots & \vdots \\ 0 & 0 & \ldots & \mathrm{sign}(c_{2^n-1}) 
    \end{pmatrix},\\ &\mathrm{sign}(A)|+\rangle^{\otimes n} = |\Tilde{\lambda}_{\mathrm{max}}\rangle.
\end{align}
QSVT is primarily used for applying polynomial functions onto matrices, hence we can only approximate the non-polynomial sign function. We provide more details on the full QSVT sequence,  as well as approximation methods for the sign function in Appendix~\ref{sec:qsvt}.

%%%%%%%%%%%%%%%%%%%%%%%%%%%%
%%%%%%%%%%%%%%%%%%%%%%%%%%%%

\subsection{Readout}

%%%%%%%%%%%%%%%%%%%%%%%%%%%%
%%%%%%%%%%%%%%%%%%%%%%%%%%%%

\subsubsection{Hypergraph state LCU preparation}

With access to $|\Tilde{\lambda}_{\mathrm{max}}\rangle$ through QSVT, the question is how to extract the signs efficiently to find the botnet. We note that the ability to read out solutions (classical information) from quantum states draws the boundary between high-performing algorithms in the $\BQP$ family and generic quantum algorithms. To perform the readout, we develop a technique based on quantum hypergraph states. The hypergraph states represent a generalization of graph states that read as \cite{Rossi_2013}
\begin{equation}\label{hypergraph_eq}
    |G\rangle = \prod_{k=1}^n\prod_{\{i_1,i_2,\ldots,i_k\}\in E} C^kZ_{i_1i_2\ldots i_k}|+\rangle^{\otimes n}.
\end{equation}
Essentially, for each k-hyperedge $\{i_1,i_2,\ldots,i_k\} \in E$, we apply a controlled-Z operation on the connected qubits. These general hypergraph states permit hyperedges of any order $k$ (for $k=1$, a single qubit $Z$ gate is applied on the corresponding qubit). Fig.~\ref{fig:Full_circuit}(d) depicts an example for $n=7$ qubits.

We can denote the set of all hypergraph states as $\mathcal{G}$. In total, $|\mathcal{G}| =\prod_{k=1}^n 2^{nC_k} = 2^{2^n-1}$, and each hypergraph state is a REW state [specified in Eq.~\eqref{REW_state}]. Since $|\Tilde{\lambda}_{\mathrm{max}}\rangle$ is also a REW state (up to some error arising from the QSVT approximation of the sign function), we know that one of our hypergraph states corresponds exactly to the bipartition that we are trying to find. In a brute force way, we can find the bipartition by simply evaluating $|\langle \Tilde{\lambda}_{\mathrm{max}}|G\rangle|^2$ (using e.g. a Hadamard test) for all hypergraph states $|G\rangle \in \mathcal{G}$ until we find $|\langle \Tilde{\lambda}_{\mathrm{max}}|G\rangle|^2 \approx 1$: the signs of this state give us the bipartition. Of course, measuring each overlap individually is costly, not to mention that we may have to trial an exponential number of hypergraph states before the correct one is found. Instead, we can exploit quantum parallelism to devise a more efficient protocol. For this, we use the linear combination of unitaries (LCU) method to create a superposition of hypergraph states, with the set of unitaries being the controlled-Z operations. Importantly, we can reduce the number of states used in the LCU, $\mathcal{G}_{\mathrm{LCU}} \subset \mathcal{G}$ if there is prior knowledge of the size of the botnet. We will provide more rationale for the exact choice of $\mathcal{G}_{\mathrm{LCU}}$ in Sec.~\ref{sec:measurement_zero}. Our hypergraph superposition state has the following form (up to normalization),
\begin{equation}
    |\psi_{\mathrm{hyp}}\rangle = \frac{1}{\sqrt{|\mathcal{G}_{\mathrm{LCU}}|}}\sum_{x=0}^{|\mathcal{G}_{\mathrm{LCU}}|-1}|x\rangle|G(x)\rangle.
\end{equation}
It is important to note that we actually do not use the full LCU protocol to create this state; we include the initial layer of Hadamards and the SELECT operator, but the final layer of Hadamards is omitted, as Fig.~\ref{fig:Full_circuit}(c) shows. This circuit gives a superposition of hypergraph states on the second register labeled by a unique bitstring on the first register. This will prove to be crucial to our method, as we will explain in the following sections.

%%%%%%%%%%%%%%%%%%%%%%%%%%%%
%%%%%%%%%%%%%%%%%%%%%%%%%%%%

\subsubsection{Projection onto signed vector}

We explained in the previous section that evaluating overlaps $|\langle \Tilde{\lambda}_{\mathrm{max}}|G\rangle|^2$ is enough to find the bipartition. By projecting onto the hypergraph superposition state, we are able to retrieve the necessary information about all these overlaps without explicitly calculating them one by one. This projector, $\hat{P}_\lambda = |\Tilde{\lambda}_{\mathrm{max}}\rangle\langle\Tilde{\lambda}_{\mathrm{max}}|$, can be retrieved from the full QSVT circuit, $\hat{U}_{\mathrm{QSVT}}$, as we detail in the following.

We know that $\mathrm{sign}(A)|+\rangle^{\otimes n} = |\Tilde{\lambda}_{\mathrm{max}}\rangle$. This matrix $\mathrm{sign}(A)$ is encoded within $\hat{U}_{\mathrm{QSVT}}$ \eqref{u_qsvt}, and we can define a state $|\psi_{\mathrm{QSVT}}\rangle = \hat{U}_{\mathrm{QSVT}}|0\rangle_a|+\rangle_s$, where we use $a=[1..n+3]$ as the set of $n+3$ ancilla qubits, and $s=[n+4..2n+3]$ denoting the set of $n$ system qubits. Hence we know if we can project using the projector $\hat{P}_{\mathrm{QSVT}} = |\psi_{\mathrm{QSVT}}\rangle \langle\psi_{\mathrm{QSVT}}|$, then measure $\boldsymbol{0}$ on the ancilla qubits, we can effectively project onto the system register with $\hat{P}_\lambda$ (since $\mathrm{tr}_a\{\hat{P}_{\mathrm{QSVT}}\} = \hat{P}_{\lambda}$). Expanding further, we have
\begin{equation}
\begin{split}
    \hat{P}_{\mathrm{QSVT}} &= |\psi_{\mathrm{QSVT}}\rangle \langle\psi_{\mathrm{QSVT}}|\\
    &= \hat{U}_{\mathrm{QSVT}}H_s|0\rangle_{as}\langle 0|_{as} H_s\hat{U}_{\mathrm{QSVT}}^\dagger \;,
\end{split}
\end{equation}
where $H_s$ represents a layer of Hadamards applied on the system register, and $|0\rangle_{as}\langle 0|_{as}$ denotes the zero projector acting on both registers. Hence this sequence for projection can be constructed as a quantum circuit with ancilla measurements. To prepare the zero projector, we note that this can be done efficiently as a simple two-term LCU \cite{alghassi2022variational},
\begin{align}
\label{eq:proj0-LCU}
|0\rangle\langle 0|^{\otimes m}= \frac{\mathds{1}_m-X^{\otimes m} \cdot C^{m-1}Z \cdot X^{\otimes m}}{2},
\end{align}
which requires a single ancilla. The full circuit to perform the projection is shown in Fig.~\ref{fig:Full_circuit}(b).

We now have the means to project onto the hypergraph superposition state, which has the following form (up to normalization):
\begin{align}\label{proj_eq}
    &(\mathds{1}\otimes\hat{P}_\lambda)|\psi_{\mathrm{hyp}}\rangle = \frac{1}{\sqrt{|\mathcal{G}_{\mathrm{LCU}}|}}\sum_{x=0}^{|\mathcal{G}_{\mathrm{LCU}}|-1}c(x)|x\rangle|\Tilde{\lambda}_{\mathrm{max}}\rangle, \\
    &c(x) = |\langle \Tilde{\lambda}_{\mathrm{max}}|G(x)\rangle|. 
\end{align}
The full circuit for this projection is shown in Fig.~\ref{fig:Full_circuit}(a).

%%%%%%%%%%%%%%%%%%%%%%%%%%%%
%%%%%%%%%%%%%%%%%%%%%%%%%%%%

\subsubsection{Measurement: zero-overlap selection}
\label{sec:measurement_zero}

The final step is to sample the top register of the circuit in Fig.~\ref{fig:Full_circuit}(a), which is the ancilla register of the hypergraph LCU. 
%%%
\begin{figure}[t!]
    \centering
    \includegraphics[width=\linewidth]{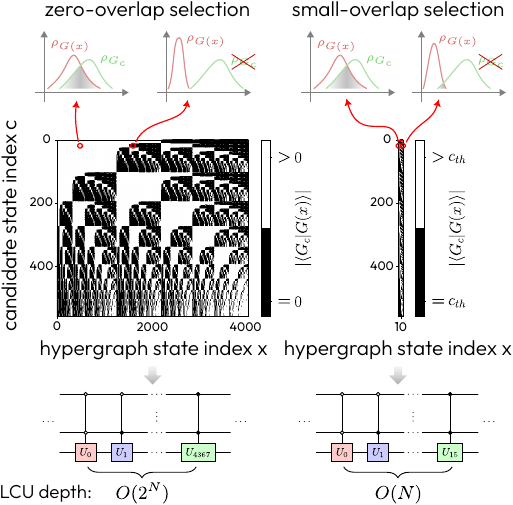}
    \caption{Comparison of the zero-overlap selection and small-overlap section methods, for $N=16$ nodes and botnet size $k=3$. We have $\binom{16}{3}=560$ candidate states, and we visualize the overlaps of these states with the states in the hypergraph LCU. For both methods, the black rectangles appear at different positions on each row, indicating that each candidate state has a unique set of zeros/small overlaps $X_2^{G_c}$. In the zero-overlap case, we select hypergraph states with botnet size $k_{\mathrm{LCU}} = \frac{16}{2}-3 = 5$, giving an LCU with $\binom{16}{5}=4368$ unitaries. In the small-overlap case, we select hypergraph states with botnet size $k_{\mathrm{LCU}} = 1$, giving an LCU with $\binom{16}{1}=16$ unitaries. Changing from zero-overlap to small-overlap selection gives us an exponential decrease in the LCU depth, which is offset by an increase in the number of circuit runs.}
    \label{fig:overlap_comparison}
\end{figure}
%%%
As Eq.~\eqref{proj_eq} shows, the probability of measuring each bitstring $x$ depends on the overlap between its corresponding hypergraph state and the signed max vector:  $p(x) \propto c(x)^2 = |\langle\Tilde{\lambda}_{\mathrm{max}}|G(x)\rangle|^2$. For a particular $|\Tilde{\lambda}_{\mathrm{max}}\rangle$, we can classify bitstrings $x$ into 2 classes,
\begin{equation}
\begin{split}
    &x \in X_1^\lambda \iff c(x)^2>0,\\
    &x \in X_2^\lambda \iff c(x)^2=0.
\end{split}
\end{equation}
Hence if we measure the first register, we are only able to measure bitstrings $x\in X_1^\lambda$. If we choose the hypergraph states used within the LCU sensibly, we will always have zeros in the probability distribution (i.e. $|X_2^\lambda|>0$). This is the property we can exploit in order to extract the botnet using just a few measurements. As for the choice of hypergraph states for the LCU, $\mathcal{G}_{\mathrm{LCU}} = \{|G(x)\rangle\} \subset \mathcal{G}$, let us assume we have a network with $N$ nodes and a botnet of size $k$. Therefore, we know that our candidate solutions $\mathcal{G}_{\mathrm{cand}} \subset \mathcal{G}$ correspond to all REW states with a botnet of size $k$, i.e. $N-k$ positive amplitudes and $k$ negative amplitudes (or vice versa, depending on convention). As for the selection of states $\mathcal{G}_{\mathrm{LCU}}$, we choose the hypergraph states with botnet of size $k_{\mathrm{LCU}} = \frac{N}{2}-k$. This selection keeps the depth of the LCU as short as possible while ensuring that each candidate solution $|G_c\rangle \in \mathcal{G}_{\mathrm{cand}}$ (including the true solution $|\Tilde{\lambda}_{\mathrm{max}}\rangle$) has a unique probability distribution and a unique set of zeros $X_2^{G_c}$,
\begin{equation}
\begin{split}
    &x \in X_1^{G_c} \iff |\langle G_c|G(x)\rangle|^2>0,\\
    &x \in X_2^{G_c} \iff |\langle G_c|G(x)\rangle|^2=0.
\end{split}
\end{equation}
For example, if we know our network has $N=16$ nodes and a botnet of size $k=3$, we have $\binom{16}{3}=560$ candidate solutions, $|\mathcal{G}_{\mathrm{cand}}|=560$. For the hypergraph LCU we choose the states with botnet of size $k_{\mathrm{LCU}} = \frac{16}{2}-3 = 5$, giving an LCU with $\binom{16}{5}=4368$ unitaries, $|\mathcal{G}_{\mathrm{LCU}}|=4368$, implementable with $\lceil \log_2(4368) \rceil = 13$ ancilla qubits. Each of the 560 candidate states has a unique set of zeros $X_2^{G_c}$, as we visualize in Fig.~\ref{fig:overlap_comparison}. In the special case where $k=\frac{N}{2}$, we select the hypergraph states with botnet size $k_{\mathrm{LCU}}=2$. If we are uncertain about the exact size of the botnet, we can extend $\mathcal{G}_{\mathrm{LCU}}$ to include hypergraph states with a range of botnet sizes. Alternatively, we can pre-calculate the size of the botnet by measuring a single overlap, as $k=\frac{N-\left(\sqrt{N}\sqrt{2^n}|\langle+^{\otimes n}|\Tilde{\lambda}_{\mathrm{max}}\rangle|\right)}{2}$.

Armed with these sets of zeros $X_2^{G_c}$ (which can be calculated easily beforehand), we use a process of sampling and elimination to find the botnet. We measure the top register of the circuit in Fig.~\ref{fig:Full_circuit}(a) in the computational basis, retrieving a bitstring $x \in X_1^\lambda$, labeling a hypergraph state $|G(x)\rangle$. With this single measurement, we can deterministically eliminate each candidate state $|G_c\rangle$ where $x\in X_2^{G_c}$; as it would have been impossible to sample $x$ if $|G_c\rangle$ is our solution state $|\Tilde{\lambda}_{\mathrm{max}}\rangle$. Returning to the example in Fig.~\ref{fig:overlap_comparison}, if bitstring $x=1000$ was measured, we would eliminate $G_c \in \mathcal{G}_{\mathrm{cand}}$ for $c \in [105,107,108,109,110,114,115,...,557,558,559]$. 

We repeat the process of measurement then elimination until one candidate state remains, this $|G_c\rangle$ is our true solution $|\Tilde{\lambda}_{\mathrm{max}}\rangle$. We simply take the signs of this state (which we know already) to retrieve the botnet. 
%%%
\begin{figure}[t!]
    \centering
    \includegraphics[width=\linewidth]{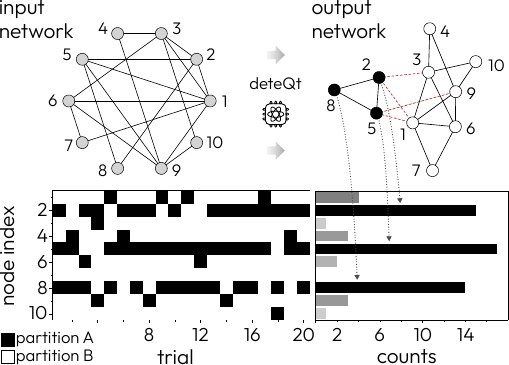}
    \caption{Example of a \emph{deteQt} run using small-overlap selection. The network considered consists of $N=10$ nodes (vertices), 18 interactions (edges), and a hidden botnet of size $k=3$. We run \emph{deteQt} 20 times (trials) reading out the estimated partitions A and B. By counting the occurrences of the single nodes in the partitions, we build the probability distribution to estimate the botnet composition (in our example, 2-5-8).}
    \label{fig:example_run_distribution}
\end{figure}
%%%

%%%%%%%%%%%%%%%%%%%%%%%%%%%%
%%%%%%%%%%%%%%%%%%%%%%%%%%%%

\subsubsection{Measurement: small-overlap selection}
\label{sec:measurement_small}

In theory, the procedure detailed above allows us to find $|\Tilde{\lambda}_{\mathrm{max}}\rangle$ with near-certainty within a logarithmic number of measurements with respect to the number of potential solutions. However, this comes with the caveat of a near-exponential depth hypergraph LCU circuit in the case when the botnet size $k$ is small, $\binom{N}{\frac{N}{2}-k}\sim \binom{N}{\frac{N}{2}}\rightarrow 2^N$ for $k\rightarrow 1$. While LCU circuits for certain groups of REW states can be simplified to have efficient representations \cite{umeano2024geometric}, this is challenging for arbitrary composition of states.

To counter this, we can venture away from selecting the hypergraph LCU states based on finding zero overlaps, $|\langle G_c|G(x)\rangle|^2=0$. To illustrate the point, rather than selecting $k_{\mathrm{LCU}} = \frac{N}{2} - k$, we now select $k_{\mathrm{LCU}} < \frac{N}{2} - k$. Each candidate solution $|G_c\rangle \in \mathcal{G}_{\mathrm{cand}}$, including the true solution $|\Tilde{\lambda}_{\mathrm{max}}\rangle$ still has a unique probability distribution, but we now separate bitstrings based on small and large overlaps, instead of zero and non-zero overlaps:
\begin{equation}
\begin{split}
    &x \in X_1^{G_c} \iff |\langle G_c|G(x)\rangle|^2>c_{\mathrm{th}},\\
    &x \in X_2^{G_c} \iff |\langle G_c|G(x)\rangle|^2=c_{\mathrm{th}}.
\end{split}
\end{equation}

This threshold overlap is defined as $c_{\mathrm{th}} = \frac{1}{\sqrt{N2^n}}|2(\frac{N}{2}-k-k_{\mathrm{LCU}})|$. This is the smallest possible overlap between a candidate state $|G_c\rangle$ with botnet size $k$ and a hypergraph LCU state $|G(x)\rangle$ with botnet size $k_{\mathrm{LCU}}$. Again, each set of small overlaps $X_2^{G_c}$, including that of our true solution $X_2^{\lambda}$, is unique to each candidate state. This allows for a significant reduction in the LCU circuit depth, as Fig.~\ref{fig:overlap_comparison} shows. We apply the same sampling and elimination process to find the botnet, with the caveat that we now have a non-zero probability of sampling $x \in X_2^{\lambda}$. Hence there is a chance of eliminating the true solution $|\Tilde{\lambda}_{\mathrm{max}}\rangle$ within this process; we cannot be 100\% sure that the last candidate state standing is indeed the true solution.

To cater for this, we repeat the protocol multiple times, before collating statistics to determine what our true botnet is. We show an example of this in Fig.~\ref{fig:example_run_distribution}. This network has $N=10$ nodes and a botnet size $k=3$, but we select $k_{\mathrm{LCU}}$ = 1. This gives us a threshold for small overlaps of $c_{\mathrm{th}} = \frac{1}{\sqrt{10\cdot16}}|2(\frac{10}{2}-3-1)| \approx 0.158$. Even though the protocol only finds the state corresponding to the true botnet 7 times out of 20 trials, the other trial solutions found have high overlap with our true solution. Counting the frequency at which each node appears in our trial solutions clearly unveils the true botnet.

\section{Protocol: scaling analysis}
\label{sec:scaling}

We now present some prospective scalings for the end-to-end protocol, extremely important to analyze as we test its ability to outperform classical analysis. We give these scalings as a function of: 1) number of nodes in the network, $N$; 2) size of the botnet, $k$; 3) degree of the polynomial approximation of the sign function, $d$. 
%%
%\begin{enumerate}
%    \item Number of nodes in the network, $N$.
%    \item Size of the botnet, $k$.
%    \item Degree of the polynomial approximation of the sign function, $d$.
%\end{enumerate}
%%
Here we assume that we are dealing with a network with a known botnet of size $k$, such that we can restrict $|\mathcal{G}_{\mathrm{LCU}}|$ as detailed previously. During this report we have written qubit counts as a function of $n=\lceil \log_2 N \rceil$. So any quantity that has a linear scaling with the number of qubits, actually has a logarithmic scaling with the number of nodes, $O(n) = O(\log N)$. 

\subsection{Number of qubits}

Analyzing Fig.~\ref{fig:Full_circuit}(a), we can see that the full circuit for the protocol uses $2n+4 + \log_2 |\mathcal{G}_{\mathrm{LCU}}|$ qubits; $2n+4 = O(\log N)$ and $|\mathcal{G}_{\mathrm{LCU}}|$ depends on whether we choose to perform readout using zero-overlap selection or small-overlap selection.

If we elect for the small-overlap method, we choose $k_{\mathrm{LCU}}$ as a small integer $a$ independent of botnet size $k$. We have $|\mathcal{G}_{\mathrm{LCU}}| = \binom{N}{k_{\mathrm{LCU}}}$, and since $a\ll N$, it follows that $|\mathcal{G}_{\mathrm{LCU}}| = \binom{N}{a} \sim O(N^a)$. We are generally free to choose $k_{\mathrm{LCU}}=1$; this gives us an exact linear scaling, $|\mathcal{G}_{\mathrm{LCU}}|= \binom{N}{1} = N$.

On the other hand, if we are searching for a botnet of size $k$ and we choose the zero-overlap method, we prepare the hypergraph LCU using circuits corresponding to states with botnet of size $k_{\mathrm{LCU}} = \frac{N}{2}-k$, and $|\mathcal{G}_{\mathrm{LCU}}| = \binom{N}{k_{\mathrm{LCU}}}$. We acknowledge that in the extreme worst case, when $k\simeq 1$ $(k_{\mathrm{LCU}} \simeq \frac{N}{2})$, $|\mathcal{G}_{\mathrm{LCU}}| \sim O(c^N)$, with $1<c<2$. However a more generic network tends to have larger communities, hence we give the scaling for these more common cases. Specifically, we take the case when $k=\frac{N}{2}-a$ $(k_{\mathrm{LCU}} = a)$, where $a$ is some integer and $a \ll N$. As in the small-overlap case, the number of states in the LCU in this situation roughly scales polynomially: $|\mathcal{G}_{\mathrm{LCU}}| = \binom{N}{a} \sim O(N^a)$. 

Overall, both approaches give us a total qubit count of $O(\log N)+O(\log N^a) \simeq \boldsymbol{O(a\log N)}$.

\subsection{Gatecounts}

For the gatecounts/circuit depth, we give scalings in terms of the number of single and two qubit gates required to implement the circuit.

Looking again at the circuit in Fig.~\ref{fig:Full_circuit}(a), we have 2 main blocks to analyze: the hypergraph LCU and the unitary $\hat{U}_\lambda$ projecting onto the signed max vector. Starting with the former, the depth of the hypergraph LCU circuit scales with the number of unitaries in the LCU, $|\mathcal{G}_{\mathrm{LCU}}| \sim O(N^a)$. Each hypergraph state is made up of a constant number of multi-qubit-controlled gates (exemplified in Fig. \ref{fig:Full_circuit}d), and an $n=\log N$-qubit controlled operator can be broken down using $O(\log N)$ single and two qubit gates \cite{rattew2023non}. Therefore, the circuit depth of the hypergraph LCU is $O(N^a \log N)$.

As for $\hat{U}_\lambda$, the circuit depth is dominated by the 2 copies of $\hat{U}_{\mathrm{QSVT}}$ (Fig.~\ref{fig:Full_circuit}b), which in turn is dominated by the $d$ repetitions of the block encoding circuit $\hat{U}_{\mathrm{BE}}$ (Fig.~\ref{fig:QSVT_implementation}). $\hat{U}_{\mathrm{BE}}$ contains 6 queries to controlled-$\hat{U}_{\mathrm{max}}$ (Fig.~\ref{fig:U_BE}, also see \cite{rattew2023non}), so the scaling depends on the method of ground state preparation. As an example, we take $\hat{U}_{\mathrm{max}}$ to be a VQE ansatz with a depth equal to the number of qubits. This gives us a $\log N$-qubit circuit with $O(\log N)$ depth: $\hat{U}_{\mathrm{max}}$ has a total of $O(\log^2 N)$ gates. As for controlled-$\hat{U}_{\mathrm{max}}$, each of the $\log^2 N$ gates can be decomposed using $O(\log N)$ single/two qubit gates, giving a circuit depth of $O(\log^3 N)$ gates for $\hat{U}_{\mathrm{BE}}$. The QSVT circuit contains $d$ repetitions of $\hat{U}_{\mathrm{BE}}$, giving $\hat{U}_{\mathrm{QSVT}}$, and hence $\hat{U}_\lambda$ a circuit depth of $O(d\log^3 N)$.

Therefore, the circuit depth for the full protocol is $\boldsymbol{O(N^a \log N)+O(d\log^3 N)}$. The term that dominates is determined by the size of the botnet, the method of ground state preparation and the desired error of the sign function approximation. 

\subsection{Circuit repetitions}

Fig.~\ref{fig:Full_circuit}(a) showcases one reason that circuit repetitions are required: only when the $n+4$ ancilla qubits are measured as $\boldsymbol{0}$ is the signed max vector projected onto the hypergraph superposition state. The number of repetitions required depends on the probability of measuring $\boldsymbol{0}$, and there are methods of boosting this probability, but in general we require $O(N)$ repetitions to perform the projection.

As for the sampling, as we see in Fig.~\ref{fig:overlap_comparison}, just a single sampled bitstring $x$ is enough to eliminate roughly half of the set of candidate solutions. This is where we see a trade-off appearing: networks with larger botnets require less states in the hypergraph LCU $\mathcal{G}_{\mathrm{LCU}}$, hence smaller circuit depths, but have a larger number of candidate states $\mathcal{G}_{\mathrm{cand}}$, which requires a larger number of measurements to complete the filtration. Dealing again with the case where $k=\frac{N}{2}-a$, we have $\mathcal{G}_{\mathrm{cand}} \sim O(c^N)$, with $1<c<2$. Taking the assumption that roughly half the set of candidate solutions are eliminated after each measurement, we require $O(\log c^N) \approx O(N)$ samples to go from the full set of candidates to a trial solution. 

For both readout methods, some repetitions of the full protocol are required to ensure the accuracy of the final solution. In the small-overlap selection case, as mentioned in Sec.~\ref{sec:measurement_small}, we repeat the protocol several times to cater for the significant chance of eliminating the true solution within a single trial. The number of trials required depends on the size of these small overlaps, $c_{\mathrm{th}} = \frac{1}{\sqrt{N2^n}}|2(\frac{N}{2}-k-k_{\mathrm{LCU}})|$. This threshold overlap is proportional to the probability of measuring a bitstring $x \in X_2^{\lambda}$, leading to incorrect elimination of the true solution. Since we take $O(N)$ samples within a single trial, the probability of reaching the end of the trial without eliminating the correct solution can be low, particularly at larger system sizes. Heuristically, we find that taking statistics from $O\left(N(\frac{N}{2}-k-k_{\mathrm{LCU}})\right)\lesssim O(N^2)$ trials is enough to mitigate for these issues.

Even in the zero-overlap selection case, we note that there is a very small error probability arising from the fact that we can only approximately apply the sign function. The signed max vector is not exactly a REW state, so there is a very small chance that we measure bitstrings that we do not expect to measure if the sign function was applied perfectly. In other words, there is a non-zero probability of measuring bitstrings $x\in X_2^\lambda$, leading to incorrect elimination of candidate states. The probability of this occurring varies from case to case, as the sign function is applied more accurately when the amplitudes of the max vector are larger; this becomes more problematic at larger system sizes when the amplitudes become smaller on average. With the $1/\sqrt{N}$ scaling in the amplitudes of a uniform state with $N$ components, we propose repeating the whole procedure $O(\sqrt{N})$ times to ensure the accuracy of our solution.

Taking the $O(N)$ repetitions to perform the zero projection, the $O(N)$ samples to find a solution from a single trial, and the number of trials needed to mitigate for erroneous candidate state elimination, which ranges from $O(\sqrt{N})$ to $O(N^2)$, the total number of circuit repetitions is upper bounded by $\boldsymbol{O(N^4)}$.

%%%%%%%%%%%%%%%%%%%%%%%%%%%%
%%%%%%%%%%%%%%%%%%%%%%%%%%%%
%%%%%%%%%%%%%%%%%%%%%%%%%%%%

\section{Applications}
\label{sec:examples}

Community detection, and more precisely botnet detection, are real--world use cases of what more largely can be referred to as the graph partitioning problem.  

One of the critical applications of community detection is anomaly detection in Intrusion Detection Systems (IDS). Network traffic can be represented as a graph, where nodes represent devices or addresses, and edges represent communication between them. Under normal conditions, the network exhibits a particular community structure, where clusters of nodes represent typical communication patterns within the network. Community detection algorithms can help identify these normal patterns by detecting dense subgraphs corresponding to routine network activity. However, when an anomaly occurs --- such as a cyber attack, a new worm propagation, or unauthorized access --- the network community structure can significantly change. When integrated into IDS, community detection methods can identify sophisticated attacks that traditional signature-based methods might miss. Analyzing the evolution of community structures can reveal stealthy threats like advanced persistent threats, which gradually expand their influence across the network over time.

Similarly, botnets --- networks of compromised devices controlled by an attacker --- often communicate with a central command-and-control (C2) server. The nodes in the botnet tend to form a community within the larger network, characterized by frequent and coordinated communication. Community detection can identify these clusters of compromised devices, enabling cybersecurity teams to isolate the botnet and cut off communication with the C2 server. 

Finally, insider threats, where individuals within an organization misuse their access to harm the organization, pose a significant security challenge. These threats can be difficult to detect because of the legitimate access insiders have to the network and its resources. These covert networks can hide within large communication networks. An insider threat might begin to interact with communities or access resources that are outside their typical community. Community detection algorithms can continuously monitor the network to identify such changes in behavior, flagging potential insider threats.

\begin{figure}
    \centering
    \includegraphics[width=\linewidth]{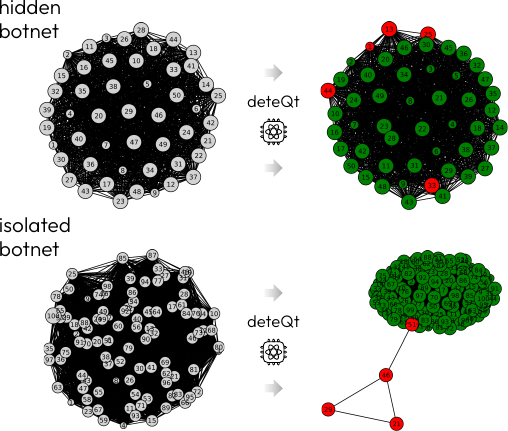}
    \caption{Example of a \emph{deteQt} run to detect botnets of different types and sizes. We show botnets that mix with the network and behave similarly to healthy nodes (hidden botnets) or botnets whose intranet is isolated from the rest of the network, limiting their exposure to the healthy nodes (isolated botnets). In the case of the hidden botnet ($k=5$), the network studied is composed of $N=50$ nodes (vertices) and 1213 interactions (edges) with a total of more than two million possible partitions. In the case of the isolated botnet ($k=4$), the network studied is composed of $N=100$ nodes and 4624 interactions, with a total of approximately four million possible partitions.}
    \label{fig:example_runs_large}
\end{figure}

In Fig.~\ref{fig:example_runs_large}, we show example applications of \emph{deteQt} to networks with non--trivial dimensionalities and structures. First, we show an example of a botnet hidden within the global network that behaves like healthy nodes, that is, having similar connection patterns. In this case, the botnet size is $k=5$, the number of nodes is $N=50$ with 1213 interactions. Using the small--overlap selection method described in Sec.~\ref{sec:measurement_small}, \emph{deteQt} is able to find the solution to the botnet out of a total of more than two million possible partitions. To do this, we needed to run the protocol approximately $10^3$ times. As a second example, we show a case of an isolated network. In the case of the isolated botnet ($k=4$), the network studied is composed of $N=100$ nodes and 4624 interactions, with a total of approximately four million possible partitions. To resolve the botnet probability distribution in this case, we needed approximately $10^4$ trials.

%%%%%%%%%%%%%%%%%%%%%%%%%%%%
%%%%%%%%%%%%%%%%%%%%%%%%%%%%
%%%%%%%%%%%%%%%%%%%%%%%%%%%%

\section{Conclusions}
\label{sec:conclusions}

In this paper, we have developed quantum computing algorithms for solving problems in complex network analysis, relevant for applications in cyber defence. As such, we introduced \emph{deteQt} --- a quantum protocol for resolving the community structure of networks and graphs, offering memory and processing advantages. Motivated by spectral methods, our approach is based on analyzing the leading eigenvector of a modularity matrix (or graph Laplacian) which contains crucial information about communities. Using the tools of ground state preparation and quantum signal processing, we prepared the corresponding real equally weighted quantum state, and proposed techniques for efficient readout of community indices based on deterministic elimination with hypergraph states. Specifically, we apply our approach to the problem of botnet detection, that is, the identification of malicious sub-networks within larger networks. As the search of suitable use-cases for early fault-tolerant quantum computers continues, we suggest that complex networks analysis represents a highly promising yet largely overlooked application area. Here, quantum detection can enhance response times to emerging threats and provide powerful tools for addressing complex attacks.
%With the clock rates of future fault-tolerant quantum computers, this advancement could bring detection times down to levels unachievable by classical methods. W

%%%%%%%%%%%%%%%%%%%%%%%%%%%%
%%%%%%%%%%%%%%%%%%%%%%%%%%%%
%%%%%%%%%%%%%%%%%%%%%%%%%%%%

\begin{acknowledgments}
Research funded by Frazer-Nash Consultancy Ltd. On behalf of the Defence Science and Technology Laboratory (Dstl) which is an executive agency of the UK Ministry of Defence providing world class expertise and delivering cutting-edge science and technology for the benefit of the nation and allies. The research supports the Autonomous Resilient Cyber Defence (ARCD) project within the Dstl Cyber Defence Enhancement programme.
%The authors acknowledge the support from the FN project SERAPIS LOT6, commissioned by dstl.
%The views expressed are those of the authors and do not reflect the official policy or position of Frazer Nash and dstl teams.
\end{acknowledgments}

%\bibliographystyle{unsrtnat_compact} % unsrtnat is loaded with revtex; introduces hyperlink doi
%\bibliography{main}
\input{main_deteQt.bbl}

%%%%%%%%%%%%%%%%%%%%%%%%%%%%
%%%%%%%%%%%%%%%%%%%%%%%%%%%%
%%%%%%%%%%%%%%%%%%%%%%%%%%%%

\appendix
\section*{Appendix}
\label{sec:appendix}

%%%%%%%%%%%%%%%%%%%%%%%%%%%%
%%%%%%%%%%%%%%%%%%%%%%%%%%%%

\subsection{Quantum Singular Value Transformation}
\label{sec:qsvt}

Block-encoding is an approach for representing a complex operator $A$ via embedding it into a unitary matrix $U$ \cite{Dong2021}. As explained in the main text, we can prepare a $(2n+2)$-qubit block-encoding unitary $\hat{U}_{\mathrm{BE}}$ which encodes a diagonal matrix $A$ containing the amplitudes of our max vector $|\lambda_{\mathrm{max}}\rangle$. This block encoding circuit involves six calls to the controlled unitary preparation circuit $\hat{U}_{\mathrm{max}}$ \cite{rattew2023non}.

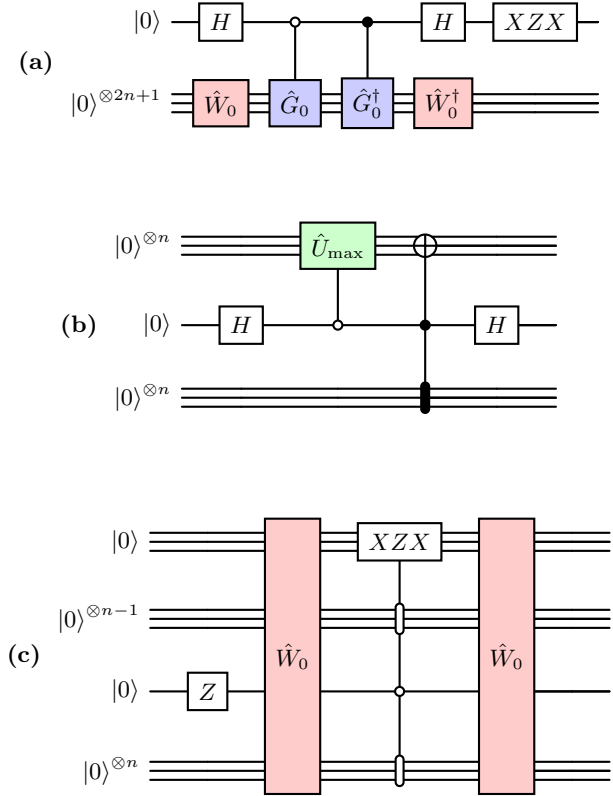
\begin{figure}
    % Panel (a)
    \begin{minipage}{0.49\textwidth}
        \textbf{(a)}
        \vspace{0.3cm}
        \begin{quantikz}[wire types={q,b},classical gap=0.12cm,column sep=0.28cm]
            \lstick{$\ket{0}$} & \gate{H} & \octrl{1} & \ctrl{1} & \gate{H} & \gate{XZX} & \qw \\
            \lstick{$\ket{0}^{\otimes 2n+1}$} & \gate[style={fill=red!20}]{\hat{W}_0} & \gate[style={fill=blue!20}]{\hat{G}_0} & \gate[style={fill=blue!20}]{\hat{G}_0^\dagger} & \gate[style={fill=red!20}]{\hat{W}_0^\dagger} & \qw & \qw \\
        \end{quantikz}
    \end{minipage}

    \vspace{0.3cm}

    % Panel (b)
    \begin{minipage}{0.49\textwidth}
        \textbf{(b)}
        \vspace{0.3cm}
        \begin{quantikz}[wire types={b,q,b},classical gap=0.12cm]
            \lstick{$\ket{0}^{\otimes n}$} &  & \gate[style={fill=green!20}]{\hat{U}_{\mathrm{max}}} & \targ{} & \qw & \qw \\
            \lstick{$\ket{0}$} & \gate{H} & \octrl{-1} & \control{} & \gate{H} & \qw \\
            \lstick{$\ket{0}^{\otimes n}$} & & & \ctrl{-2} & \qw & \qw \\
        \end{quantikz}
    \end{minipage}
    \hfill
    
    \vspace{0.4cm}
    % Panel (c)
    \begin{minipage}[t]{0.49\textwidth}
        \textbf{(c)}
        \vspace{0.3cm}
        \begin{quantikz}[wire types={b,b,q,b},classical gap=0.12cm]
            \lstick{$\ket{0}$} &  & \gate[wires=4,style={fill=red!20}]{\hat{W}_0} & \gate{XZX} & \gate[wires=4,style={fill=red!20}]{\hat{W}_0} & \qw & \qw \\
            \lstick{$\ket{0}^{\otimes n-1}$} &  & & \ocontrol{} & & \qw & \qw \\
            \lstick{$\ket{0}$} & \gate{Z} & & \ocontrol{} & & \qw & \qw \\
            \lstick{$\ket{0}^{\otimes n}$} & & & \octrl{-3} & & \qw & \qw \\
        \end{quantikz}
    \end{minipage}
    \caption{(a) Full circuit for $\hat{U}_{\mathrm{BE}}$, the diagonal block encoding of the amplitudes of $|\lambda_{\mathrm{max}}\rangle$. The block $XZX$ represents the sequences of single qubit gates $X, Z, X$. The circuits for the components $\hat{W}_0$ and $\hat{G}_0$ are given in (b) and (c) respectively. See \cite{rattew2023non} for the full details of this block encoding procedure.} 
    \label{fig:U_BE}
\end{figure}

Once we have represented operators in the block-encoded form [Eq.~\eqref{Block_enc_eq}], we can proceed with information processing based on the unitary operators. The beauty of such a representation is in the ability to manipulate eigenvalues of associated operators by controlled actions conditioned on just one qubit. This corresponds to quantum signal processing (QSP) and quantum singular value transformation (QSVT) techniques. In their essence, these techniques represent polynomial transformations of eigenvalues $\lambda \mapsto P(\lambda)$, where $P(x)$ is a polynomial approximation of some selected function $f(x) \approx P(x)$. QSVT is the approach used when transformation is applied to arbitrary operators (matrices) \cite{Gilyen2019,Martyn2021}. 
%An alternative approach to matrix inversion is through applying an approximation to the inverse function, representing it as Chebyshev or Taylor series.

Generally, quantum signal processing designs polynomials of any variable $x$ by embedding it with single-qubit rotations, based on the additional phases $\boldsymbol{\phi}$ that define the type of polynomial \cite{Low2017}. This polynomial transformation is achieved through a sequence of unitary operations \cite{Martyn2021}:
\begin{align}
U(\boldsymbol{\phi},x) = e^{i\phi_0 Z} \prod_{k=1}^{d} \left( W(x) e^{i\phi_k Z} W(x)^\dagger \right).
\end{align}

$W(x)$ is called the signal rotation operator which encodes the scalar $x$, and $\boldsymbol{\phi}$ is a vector of phase angles. This sequence with the correct sequence of angles performs the transformation $x \mapsto P(x)$, where $P(x) = \langle0|U(\boldsymbol{\phi},x)|0\rangle$. 

\begin{figure}[h!]
    \centering
    \includegraphics[width=\linewidth]{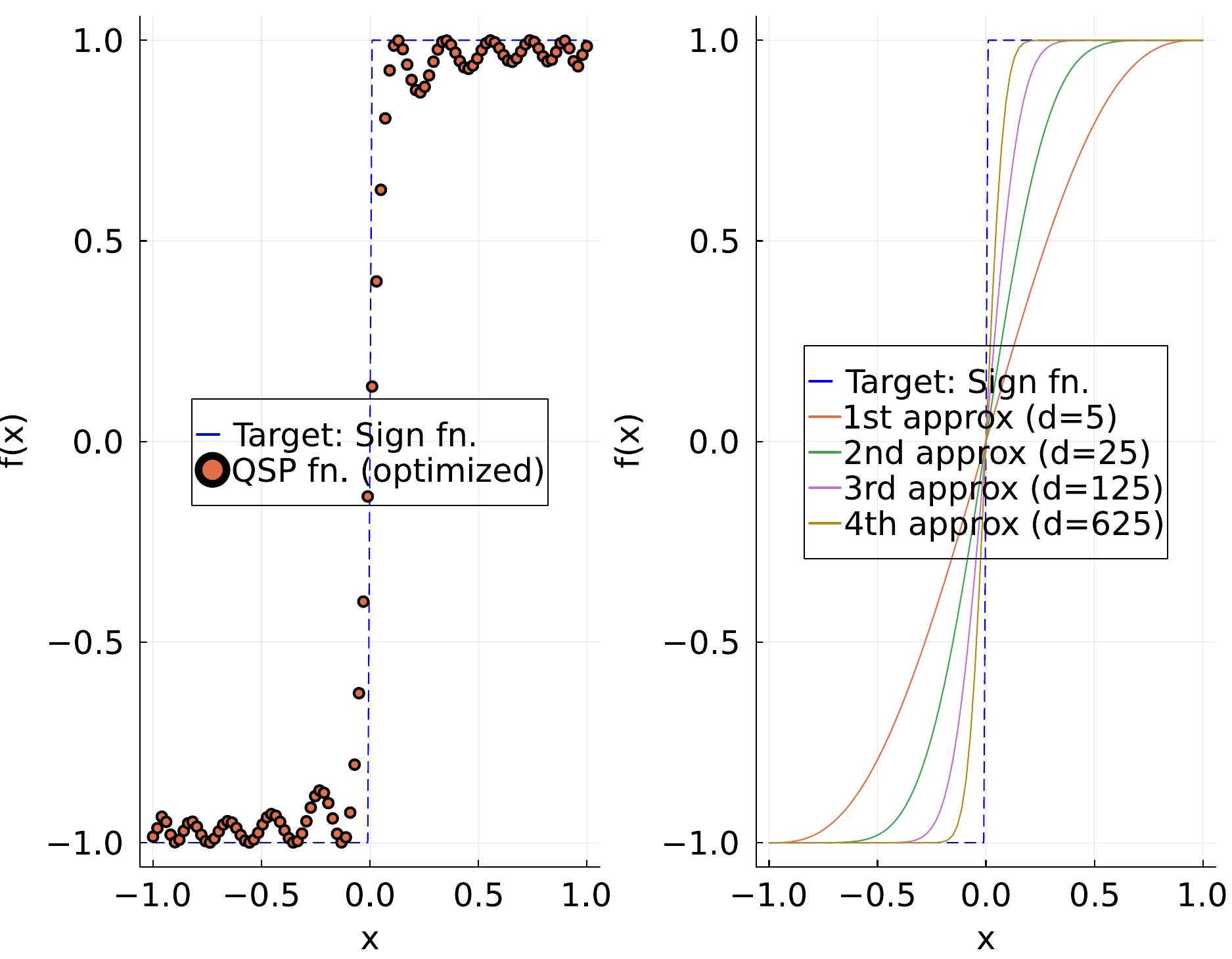}
    \caption{QSP approximation of the sign function using angles generated by optimisation (left) and recursive methods (right). For the optimisation, $d = 29$ angles were used to achieve a decent approximation, while for the recursive method we plot the 1st, 2nd, 3rd and 4th order approximations, corresponding to $d={5, 25, 125, 625}$.}
    \label{fig:QSP_sign_fn}
\end{figure}

\begin{figure*}[h!]
   \centering
   \begin{quantikz}[wire types={q,b,b}, classical gap=0.08cm, row sep=0.2cm, column sep=0.3cm]
   \lstick{$\ket{0}$} & \targ{}\gategroup[3,steps=3,style={dashed,rounded corners,fill=blue!20, inner xsep=1.5pt},background,label style={label position=below,anchor=north,yshift=-0.2cm}]{{$\Pi_{\phi_1}$}} & \gate{R_z(2\phi_1)} & \targ{} && \targ{}\gategroup[3,steps=3,style={dashed,rounded corners,fill=blue!20, inner xsep=1.5pt},background,label style={label position=below,anchor=north,yshift=-0.2cm}]{{$\Pi_{\phi_2}$}} & \gate{R_z(2\phi_2)} & \targ{} &&\cdots & \meter{} \\
   \lstick{$\ket{0}^{\otimes n+2}$} & \ocontrol{} && \ocontrol{} & \gate[wires=2,style={fill=orange!20}]{\hat{U}_{\mathrm{BE}}} &\ocontrol{} && \ocontrol{} & \gate[wires=2,style={fill=orange!20}]{\hat{U}_{\mathrm{BE}}^\dagger}&\cdots & \meter{}\\
   \lstick{$\ket{\psi}$} & \octrl{-2} && \octrl{-2} && \octrl{-2} && \octrl{-2} &&\cdots & \rstick{$\mathrm{sign}(A)\ket{\psi}$}
   \end{quantikz}
   \caption{Snippet of the QSVT sequence $\hat{U}_{\mathrm{QSVT}}$ for the transformation $A\rightarrow \mathrm{sign}(A)$. If $\mathbf{0}$ is measured on the ancilla qubits, the transformed matrix $\mathrm{sign}(A)$ is applied onto the system register $\ket{\psi}$. The accuracy of the application of the sign function depends on the choice of angles $\boldsymbol{\phi}$.}
   \label{fig:QSVT_implementation}
\end{figure*}
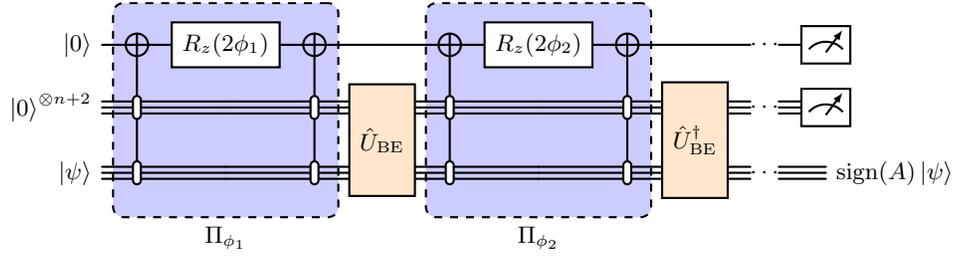

As noted in the main text, QSP is optimized for polynomial functions of fixed degree. Yet, we can still find angles that can approximate the non-polynomial sign function. For example, we can opt for brute force optimization, or alternatively, recursive methods using an analytically found set of angles is also effective \cite{mizuta2024recursive}. We show a comparison between the two methods in Fig.~\ref{fig:QSP_sign_fn}. QSP with optimization reaches a good approximation with fewer angles (and hence shorter circuit depth), but recursive methods provide smoother functions and give better convergence to the exact sign function at higher orders.
%%%
%%%
QSVT, a generalization of QSP, specifically targets the singular values of the matrix $A$. It allows the implementation of a desired polynomial transformation $P(A)$ on the singular values of $A$ using a sequence of block-encoded operations and phase rotations. The QSVT sequence is made up of block encoding unitaries $\hat{U}_{\mathrm{BE}}$ interleaved with projector-controlled phase (PCP) gates $\Pi_\phi$. Since our block encoding uses $2n+2$ qubits, we require the PCP gates to apply $e^{i\phi}$ onto the subspace corresponding to the $n$ system qubits, and $e^{-i\phi}$ to the $(n+2)$-dimensional ancilla subspace. We can implement this PCP gate using an additional ancilla, and we show part of the full QSVT sequence in Fig.~\ref{fig:QSVT_implementation}.

The full QSVT sequence acts on the singular values of $A$ such that
\begin{align}
P(A) = U P(\Sigma) V^\dagger, 
\end{align}
where $A = U \Sigma V^\dagger$ is the singular value decomposition of $A$, and $\Sigma$ is the diagonal matrix of singular values.

%%%%%%%%%%%%%%%%%%%%%%%%%%%%
%%%%%%%%%%%%%%%%%%%%%%%%%%%%
%%%%%%%%%%%%%%%%%%%%%%%%%%%%

\end{document}